\documentclass[
 aip,
 amsmath,amssymb,
 preprint,%
]{article}

\usepackage{graphicx}
\usepackage{subfig}
\usepackage{dcolumn}
\usepackage{bm}
\usepackage{amsmath}
\usepackage{amssymb}
\usepackage{amsfonts}
\usepackage{amsbsy}
\usepackage{eucal}
\usepackage[T1]{fontenc}
\usepackage{multirow}
\usepackage{txfonts}
\usepackage{color}
\usepackage[normalem]{ulem}
\usepackage{authblk}

\usepackage{lineno,hyperref}
\modulolinenumbers[5]

\renewcommand{\u}{\mathbf{u}}
\newcommand{\x}{\bm{x}}

\renewcommand{\v}{\mathbf{v}}

\newcommand{\w}{\mathbf{w}}
\newcommand{\jb}{\mathbf{j}}

\newcommand{\Ubsf}{\boldsymbol{\mathsf{U}}}
\newcommand{\phib}{{\boldsymbol{\phi}}}

\newcommand{\Phib}{{\boldsymbol{\Phi}}}

\newcommand{\Xc}{\mathcal{X}}
\newcommand{\Jc}{\mathcal{J}}

\newcommand{\Nbb}{\mathbb{N}}

\newcommand{\Rbb}{\mathbb{R}}

\newcommand{\dd}{\mathrm{d}}

\DeclareSymbolFontAlphabet{\mathcal}{symbols}

\begin{document}


\title{Sparse Low Rank Approximation of Potential Energy Surfaces with Applications in Estimation of Anharmonic Zero Point Energies and Frequencies}

\author[1]{P. Rai \thanks{pmrai@sandia.gov}}
\author[1]{K. Sargsyan}
\author[1]{H. Najm}
\author[2]{S. Hirata}
\affil[1]{Sandia National Laboratories, California, USA}
\affil[2]{Department of Chemistry, University of Illinois at Urbana-Champaign, Urbana, Illinois 61801, USA}


%

%
%

\date{}
\maketitle

\newcommand{\slugmaster}{%
\slugger{juq}{xxxx}{xx}{x}{x--x}}

%

\begin{abstract}
We propose a method that exploits sparse representation of potential energy
surfaces (PES) on a polynomial basis set selected by \textit{compressed
sensing}. The method is useful for studies involving large numbers of PES
evaluations, such as the search for local minima, transition states, or integration.
We apply this method for estimating zero point energies and frequencies of
molecules using a three step approach. In the first step, we interpret the PES
as a \textit{sparse tensor} on polynomial basis and determine its entries by a
compressed sensing based algorithm using only a few PES evaluations. Then, we
implement a rank reduction strategy to compress this tensor in a suitable
low-rank canonical tensor format using standard tensor compression tools. This
allows representing a high dimensional PES as a \textit{small} sum of products
of one dimensional functions. Finally, a low dimensional Gauss-Hermite
quadrature rule is used to integrate the product of sparse canonical low-rank
representation of PES and Green's function in the second-order diagrammatic
vibrational many-body Green's function theory (XVH2) for estimation of
zero-point energies and frequencies.  Numerical tests on molecules considered in
this work suggest a more efficient scaling of computational cost with molecular
size as compared to other methods.
\end{abstract}




\section{Introduction}
\label{sec:1}
Electronic structure calculations have been developed as a powerful tool that is
used in several fields including chemical sciences and biochemistry, as well as
material and energy sciences. In \textit{ab initio} electronic structure
calculations, for instance, computation (and often storage) of six-dimensional
integrals involving two-body (i.e., Coulomb) interactions (``two-electron
integrals") is necessary, creating a severe bottleneck of such calculations for
larger molecules \cite{Almlof:1982}. In quantum dynamics, accurate estimation of
energy and vibrational frequencies of molecules requires integration of
functions whose dimensionality increases linearly with $a$, the number of nuclei
in the molecule. For example, the potential energy surfaces (PESs) are often a
part of the integrand, and their dimensionality increases as $(3a - 6)$, where $a$ is the number of atoms. Thus
efficient ways to approximate and integrate high dimensional PESs that
exploit their special structure, if it exists, are needed.

Numerical approximation or integration of these PESs, in practice, can be carried
out via sampling techniques with the function as a black-box.  For example, one
probes the PES at different configuration of atoms in a molecule with standard
quantum chemistry software packages such as NWChem \cite{nwchem}. Many methods to
represent a PES using a set of energy data points exist. Some of the black-box
fitting methods include splines \cite{bowman1986binitio,chapman1983theoretical},
modified-Shepard interpolation \cite{ischtwan1994molecular}, interpolating
moving least squares \cite{maisuradze2003interpolating, guo2004interpolating},
neural networks \cite{sumpter1992potential, blank1995neural,brown1996combining,
prudente1998fitting,gassner1998representation,lorenz2004representing}, and
reproducing kernel Hilbert space \cite{hollebeek1999constructing}. These
methods, although efficient with PES approximation of smaller systems, do not
usually scale well with system size and may suffer with severe computational
constraints. Also, the functional form of the approximation may not render it in
a way that is easy to integrate or employ further in a given computational pipeline.

Methods for accurate PES approximation of bigger molecules with relatively few
PES evaluations are needed. Mathematically, for high dimensional functions, application
of standard approximation approaches is often not sufficient due to the curse
of dimensionality, i.e. when the required computational effort increases
exponentially with dimension.  One usually employs a class of methods that
exploit specific structures of high dimensional functions, such as smoothness or
sparsity (see chapter 1 of \cite{RAI14} for a brief survey). In this work, we
first exploit the sparsity property of the PES using \textit{compressed
sensing} \cite{CAN06, CAN06b} methods from the signal processing community. The mathematical theory of
compressed sensing is well-developed and is being used in many
scientific applications. Some of the experimental applications of compressed sensing
include multidimensional nuclear magnetic resonance \cite{ANIE11(1),ANIE11(2)},
 super-resolution microscopy \cite{ZHU12}
and other applications in spectroscopy and beyond \cite{GROSS10,SHABANI11,
 SANDERS12,AUGUST13,XU14}.
Compressed sensing is also becoming a method of choice for
computational applications \cite{ANDRADE12,ALMEIDA12, SCHAEFFER13, NELSON13}.
For example, in \cite{SANDERS15}, compressed sensing is used to
reduce the amount of computation in numerical simulations for molecular vibrations.

Sparse approximation using compressed sensing relies on the fact that a good approximation of the PES
can be obtained by representing it as a linear combination of only a few basis
functions chosen from a well-constructed set of basis functions \cite{BLA11,DOO11}. Existence of sparsity structure can be attributed to the fact that the function is not \textit{equally coupled} in all dimensions and hence only a few basis functions are important for an accurate representation. Here, the basis
set consists of tensor products of orthogonal polynomials, and
its subset is obtained with a constraint imposed on the total order of these multivariate
polynomials.  The approximation obtained can then be interpreted as a
\textit{sparse tensor}. We then exploit the \textit{low rank structure}
\cite{Hackbusch:2012,Khoromskij:2012,Grasedyck:2013} in a subsequent step by
applying a rank reduction strategy on the sparse tensor.  Here, a high rank
representation of a PES is compressed as a \textit{small} sum of products of
low dimensional functions \cite{Khoromskaia:2015,benedikt2011tensor}. Finally, these
low dimensional integrals are integrated using a Gauss-Hermite quadrature rule.
The proposed approach thus presents a synthesis of three ideas: sparse
approximation, low rank compression and quadrature for the estimation of zero point energies and frequencies in XVH2. This approach is efficient, and has a
strong potential for scalability with molecular size. Note that methods that explicitly make use of low rank structure of the PES have recently been proposed \cite{RAI17,Rauhut}.

The outline of the paper is as follows. In Section~\ref{sec:sparse}, we discuss the tensor interpretation of functions and its approximation using least-squares with sparse regularization via compressed sensing. Then, in Section~\ref{sec:low_rank}, we present a general method for compressing functions represented on a tensor product basis and apply it for integrating high dimensional functions using separated integration. In Section~\ref{sec:quantumchem}, we recall formulations in quantum chemistry that lead to first and second order
corrections to the zero point energy and anharmonic molecular vibrations. In Section~\ref{sec:applications}, we
illustrate the application of the proposed method on four different molecules with increasing dimensionality of PES. Finally, we conclude with perspectives in
Section~\ref{sec:conclusion}.

\section{Sparse Approximation of Potential Energy Surfaces}
\label{sec:sparse}
Very often in scientific discovery and applications, one may collect a large amount of data but only a subset of it may be relevant to study the problem at hand. The difficulty however is that one does not know \textit{a priori} where the useful information can be found and how much information is sufficient. In approximation of PES of a system, one is required to evaluate the surface as many times as the underlying model assumption or the numerical scheme needs to provide a closed form solution. For example, consider the the representation of PES as a multivariate function $u(\x)$ expanded on a multidimensional tensor product basis as
\begin{eqnarray}
u(\x) \approx \tilde{u}(\x) = \sum_{j_1=0}^{p_1}\cdots\sum_{j_m=0}^{p_m} v_{j_1,\ldots,j_m} \phi^{(1)}_{j_1}(x_1)\cdots\phi^{(m)}_{j_m}(x_m),
\label{functional_multivariate}
\end{eqnarray}
where $\phi^{(i)}_{j_i}(x_i)$ is the $j_i$th basis function in the $i$th
coordinate, $x_i$. Here, the total number of coefficients of the
multidimensional basis needed to characterize $\tilde{u}(\x)$ is given by $P=
\prod_{k=1}^m p_k$. For the sake of simplicity, let us choose the same number
of basis functions, $p$, in each dimension such that $P=p^m$. For a high dimensional PES, $m$ is large,
and the exponential increase in bases terms is referred to as the
\textit{curse of dimensionality}. For example, a least squares based approximation of
$u(\x)$ will require at least $P$ evaluations of $u(\x)$ which, in general,  may
not be feasible. However, if we are only interested in an accurate estimate of
integration of $u(\x)$ over a domain $\Omega$, as in this work (and not, for example, in point-wise
accuracy of the surface), then we can drastically reduce the number of required evaluations of $u(\x)$.

The mathematical assumption that enables a good approximation with only a few evaluations of $u(\x)$ is that of all coefficients $v_{j_1,\ldots,j_m}$ in \eqref{functional_multivariate} only a few are nonzero. Under this assumption, $u(\x)$ is said to be \textit{sparse} on multidimensional bases $\phi^{(1)}_{j_1}(x_1)\cdots\phi^{(m)}_{j_m}(x_m)$.
The problem then reduces to finding the nonzero coefficients such that the number of evaluations of $u(\x)$ required is proportional to only the number of nonzero coefficients. In order to achieve this objective, we use a two-fold approach. First, we choose $\phi^{(i)}_{j_i}(x_i)$  as orthogonal polynomials and \textit{a priori} reduce the number of multidimensional bases in \eqref{functional_multivariate} based on their \textit{total degree} (to be defined below). Second, we further take advantage of sparsity of $u(\x)$ in its representation in the reduced basis set by using techniques from compressed sensing. Both approaches are detailed in the following subsections.

\subsection{A priori reduction of basis set}
In this section, we propose an a priori reduction of the tensor product basis set for the representation of potential energy functions under the assumption that they admit limited degree of high-order interactions. 
It is well known that, for smooth functions, polynomials are a natural choice of basis functions for functional representation. Let us denote the set of multidimensional polynomials with degree per-dimension $p$ as

\begin{equation}
\mathbb{Q}_p =\left\{\prod_{i=1}^m \phi^{(i)}_{j_{i}}(x_i): \mathbf{j}\in\Nbb_0^{m}, \vert \mathbf{j} \vert_{\infty}:=\underset{i\in\{1\ldots m\}}{\mathrm{max}}j_i\leq p\right\},
\end{equation}
where $\phi^{(i)}_{j_{i}}(x_i)$ is a polynomial of degree $j_i$ and $\mathbf{j}$ is a multi-index $(j_1,\ldots,j_m)$ in the set of multi-indices $\Jc = \times_{i=1}^m\{0,\ldots,p\}$. The total number of basis functions for a given $p$ in this set is given by $(p+1)^m$. For example, in \eqref{functional_multivariate}, $u(\x)$ is represented on the basis in $\mathbb{Q}_p$ when $p_1=\ldots=p_m = p$.
Another alternative is the set of multidimensional polynomials of
total degree $p$ defined by:
\begin{equation}
\mathbb{P}_p=\left\{\prod_{i=1}^m \phi^{(i)}_{j_i}(x_i): \mathbf{j} \in\Nbb_0^{m},\vert\mathbf{j}\vert_1:=\sum_{i=1}^m j_i\leq p\right\}.
\end{equation}
Here the total number of basis functions for a given $p$ is given by $P = \frac{(m+p)!}{m!p!}$.

\begin{figure}[htb!]
\includegraphics[scale=0.90]{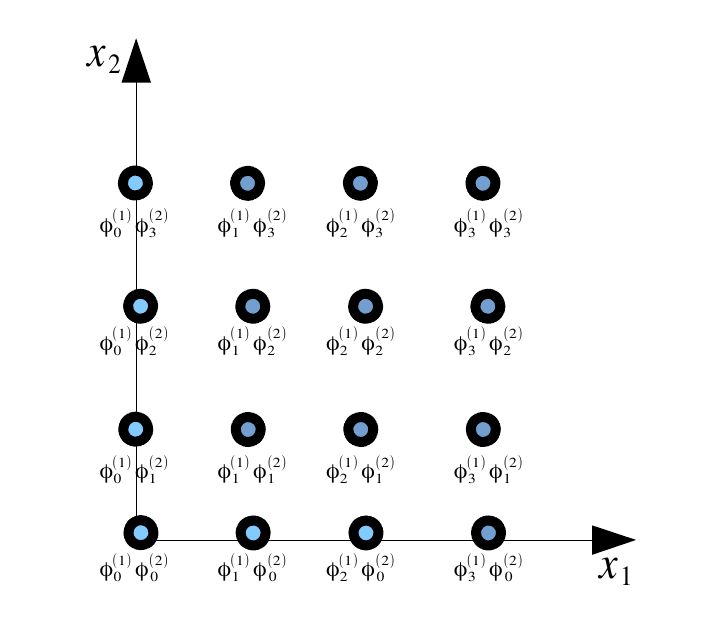}
\includegraphics[scale=0.90]{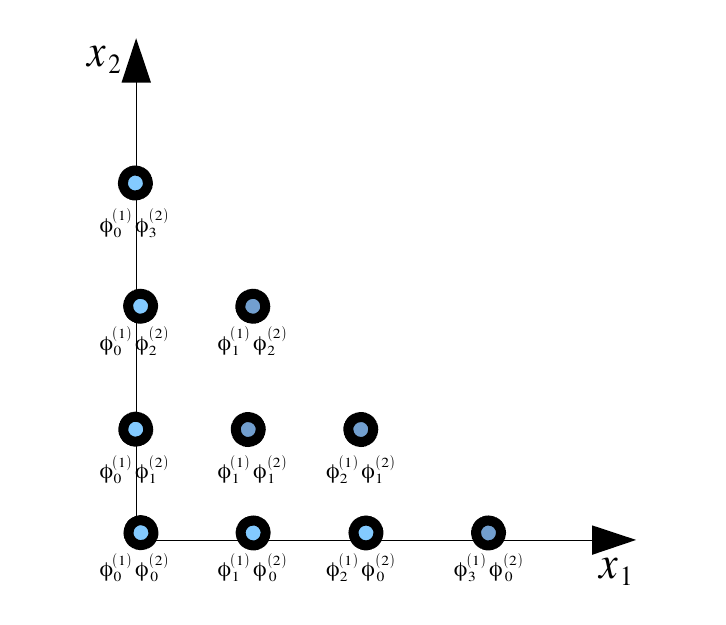}
\caption{Illustration of basis functions in $\mathbb{Q}_p$ (left) and $\mathbb{P}_p$ (right) for degree $p=3$ and dimension $m=2$. The number of basis function in $\mathbb{Q}_p$ is $(3+1)^2 = 16$ whereas in $\mathbb{P}_p$ is $\frac{(3+2)!}{3!2!} = 10.$}
\label{fig:approx_spaces}       
\end{figure}

In Figure \ref{fig:approx_spaces}, we illustrate the set of basis functions in $\mathbb{Q}_p$ and $\mathbb{P}_p$ for $p=3$ and $m=2$. It can be clearly seen that $\mathbb{P}_p$ is a subset of $\mathbb{Q}_p$, albeit the number of elements in $\mathbb{P}_p$ still grows fast with dimension. The first step in this work is to reduce the number of basis functions for representation of $u(\x)$ by choosing \textit{a priori} the basis set $\mathbb{P}_p$ (and not $\mathbb{Q}_p$). Thus we are explicitly choosing the coefficients corresponding to those basis function in $\mathbb{Q}_p$ as zero that are not present in ${\mathbb{P}_p(\Xc)}$. We can thus write the approximation of $u(\x)$ as
\begin{equation}
u(\x) \approx \tilde{u}(\x) =  \sum_{\jb\in \tilde{\Jc}} v_\jb\phi_\jb(\x),
\label{reducspace}
\end{equation}
where $\mathbf{j}\in \tilde{\Jc}$ such that $\vert \mathbf{j}\vert_1 \leq p$ and $v_{\jb}$ is real coefficient on the basis $\phi_\jb = \prod_{i=1}^m\phi_{j_i}^{(i)}$.
In the next step, we further reduce the number of basis functions with non-zero coefficients in \eqref{reducspace} using methods based on compressed sensing.

\subsection{Sparse approximation using Compressed Sensing}

\begin{figure}[htb!]
\includegraphics[scale=0.90]{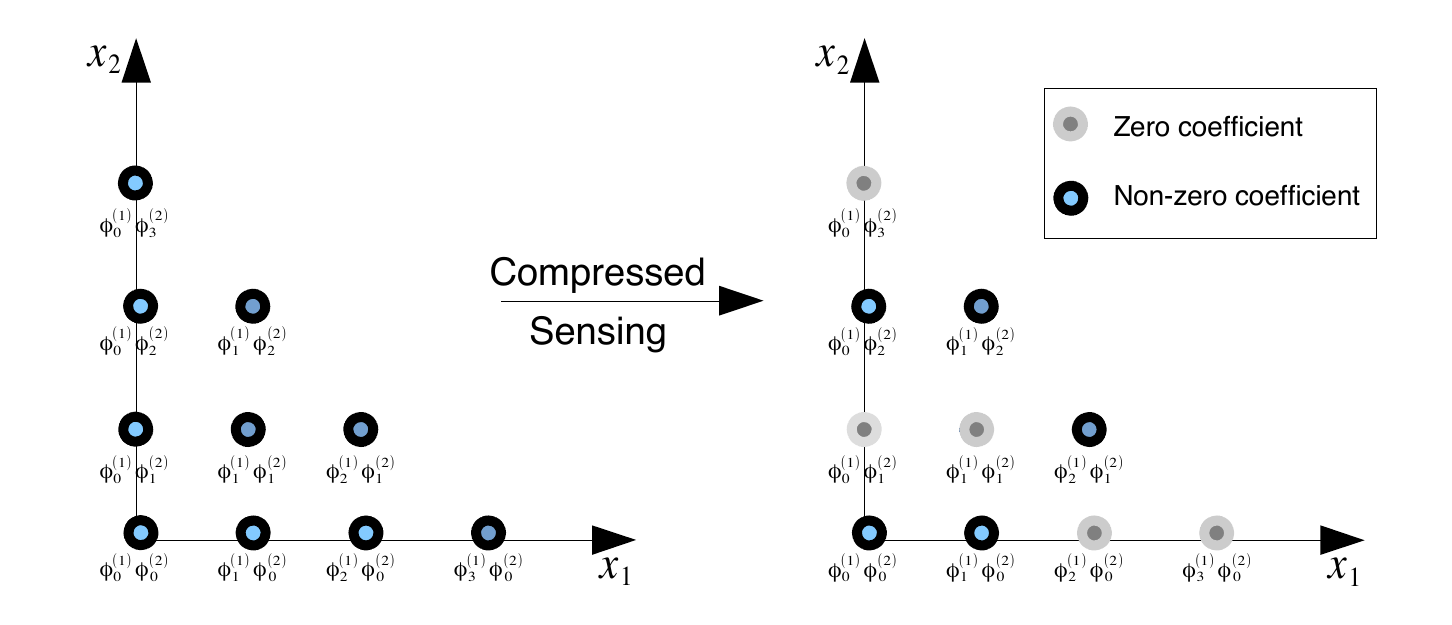}
\caption{Illustration of application of compressed sensing based selection of basis functions with nonzero coefficients (right) from the basis set $\mathbb{P}_2$ (left).}
\label{fig:cs_illustration}       
\end{figure}

Let us represent Eq. \eqref{reducspace} as a linear system of equations
\begin{equation}
\Phib\v=\u,
\label{linsys1}
\end{equation}
where $\u \in \Rbb^S$ is a vector of $S$ evaluations of $u(\x)$ on
$\mathbf{x}^s, s=1,\ldots,S$ realizations of $\x$ and $\Phib \in \Rbb^{S\times
P}$ is the matrix whose row elements are basis functions $\phi_{\jb}$ evaluated at
$\mathbf{x}^s$ i.e. $\Phib_{s\ell}=\phi_{{\jb}_\ell}(\mathbf{x}^s)$, where $\ell=1,\ldots,P$ indexes an ordering of 
$\mathbf{j}_\ell\in \tilde{\Jc}$.
One can obtain the vector of coefficients $\v \in \Rbb^{P}$ using least squares by
\begin{equation}
\v = (\Phib^{T}\Phib)^{-1}\Phib^{T}\u,
\label{linsys2}
\end{equation}
where, for a well-defined matrix inversion, one needs at least as many
independent evaluations of $u(\x)$ as the number of basis functions i.e. $S \ge
P$. In the absence of sufficient number of evaluations i.e. $S < P$,
\eqref{linsys1} is an underdetermined system with an infinite number of
solutions. However, if it is known that $\v$ is \textit{sparse}, meaning that it
has very few nonzero components, we can obtain a good approximation of $\v$ with
$S\ll P$ using compressed sensing. Equivalently, compressed sensing endeavors to
find a sufficiently accurate representation of $u(\x)$ by only considering a
finite subset of basis functions:
\begin{equation}
u(\x)\approx \tilde{u}_n(\x)=\sum_{\jb \in \tilde{\Jc}_n}v_\jb\phi_\jb(\x).
\label{lasso}
\end{equation}
Here $\tilde{\Jc}_n\subset \tilde{\Jc}$ such that $\tilde{u}_n(\x)$ is represented on only $n\ll P$ basis functions. The $n$ nonzero components of $\v$ are obtained by formulating the following optimization problem
\begin{equation}
\underset{\Vert \v\Vert_0=n}{\mathrm{min}}\Vert \Phib\v-\u\Vert_2^2,
\label{sparse_zero}
\end{equation}
where $\Vert \cdot \Vert_0$ is the zero-norm of a vector and simply counts the number of nonzero components.  The above problem is non-convex and we instead solve
\begin{equation}
\mathrm{min}\;\Vert \Phib\v-\u\Vert_2^2+\lambda\Vert \v \Vert_1,
\label{sparse_one}
\end{equation}
where $\Vert \cdot\Vert_1$ is the $\ell_1$ norm i.e. sum of absolute values of
components of a vector and $\lambda$ is a regularization coefficient. Problem
\eqref{sparse_one} is a convex optimization problem for which several methods
are available \cite{Bach12}. If $u(\x)$ admits an accurate sparse approximation,
then, under some additional conditions, solving \eqref{sparse_one} gives a
sparse solution $\v$. This is the basis of the compressed sensing approach that
made a breakthrough in signal processing more than a decade ago
\cite{CAN06,CAN06b,DONOHO06}. In this work, we use a Bayesian Compressed Sensing
\cite{SARGSYAN14} algorithm, which has is implemented in the UQTk
software~\cite{DEBUSSCHERE04}, for approximately solving \eqref{sparse_one}.  We
illustrate the application of compressed sensing in Figure
\ref{fig:cs_illustration} where we (symbolically) retain only a few basis
functions from the set basis set $\mathbb{P}_3$ with nonzero coefficients in the
\textit{sparse} approximation of $u(\x)$.

Thus by combining the two methods explained in this section, we can greatly
reduce the computational expense required to estimate a good approximation of
$u(\x)$. In high dimensions however, the number of retained basis functions $n$
can still be large and, depending on the application (ex. integration), one may
be interested in a more compact representation of $\tilde{u}_n(\x)$. In the
following section, we detail our approach for integrating a compressed version
of $\tilde{u}_n(\x)$ using separated integration.

\section{Separated Integration of Sparse Approximation}
\label{sec:low_rank}
Let us consider the following integration problem
\begin{eqnarray}
I[u] = \int\limits_{-\infty}^{+\infty} u(\x) \rho(\x) \dd\x,
\label{eq:integ}
\end{eqnarray}
where $\rho(\x)$ is a non-negative weight function that is integrable and multiplicatively separable, i.e.,
\begin{eqnarray}
\rho(\x) &\ge& 0, \label{condition1}\\
\int\limits_{-\infty}^{+\infty} \rho(\x)\dd\x &<& \infty, \label{condition2}
\end{eqnarray}
and
\begin{eqnarray}
\rho(\bm{x}) &=& \prod_{i=1}^m \rho^{(i)}(x_i).\label{condition3}
\end{eqnarray}
An approximation of \eqref{eq:integ} can be written as
\begin{eqnarray}
I[u]\approx I[\tilde{u}_n] = \int\limits_{-\infty}^{+\infty} \tilde{u}_n(\x) \rho(\x) \dd\x.
\label{eq:cs_integ}
\end{eqnarray}
Substituting the sparse approximation of $\tilde{u}_n(\x)$ from \eqref{lasso} into \eqref{eq:cs_integ}, we get
\begin{equation}
I[\tilde{u}_n] = \sum_{\jb \in \tilde{\Jc}_n}v_\jb \left(\prod_{i=1}^{m} \int\limits_{-\infty}^{+\infty}\phi^{(i)}_{j_i}\rho^{(i)}(x_i)dx_i\right),
\label{eq:cs_sep_int}
\end{equation}
where the integrands in \eqref{eq:cs_sep_int} are one dimensional polynomial
functions and can be easily evaluated using standard quadrature rules. However,
as will be seen in the second order corrections in XVH2 in Section
\ref{sec:XVH2integrals}, even if $n$ is only moderately large, the number of
quadrature integrations required can increase rapidly. In order to improve
computation efficiency, we can reduce the number of separated terms in $u_n(\x)$,
i.e. separation rank, considerably for a small loss of accuracy. We thus strive
to find approximations of the form

\begin{equation}
\tilde{u}_n(\x)\approx \tilde{u}_r(\x) = \sum_{k=1}^{r\ll n} \alpha_k w_k(\x),\; \alpha_k\in \Rbb,
\label{sparse2lr}
\end{equation}
where
\begin{equation}
w_k(\x) = \prod_{i=1}^m w^{(i)}_k(x_i),\:\textrm { and }\: w^{(i)}_k(x_i) = \sum_{j=1}^p w^{(i)}_{k,j}\phi_j^{(i)}(x_i)
\label{sparse2lrb}
\end{equation}
and $r$ is the separation rank. The univariate functions $w_k^{(i)}(x_i)$, characterized by coefficient vectors
$\w^{(i)}_k = (w^{(i)}_{k,1},\ldots,w^{(i)}_{k,p})$, are represented as
expansions on bases $\phib^{(i)} =
(\phi_1^{(i)}(x_i),\ldots,\phi_p^{(i)}(x_i))$. Note that $\w^{(i)}_k$ and hence $w_k^{(i)}(x_i)$ need not be sparse. Finding low rank approximations
of the form \eqref{sparse2lr} thus reduces to finding coefficient vectors
$\w^{(i)}_k,\; 1\leq i\leq m, 1\leq k\leq r$. The scalar coefficient $\alpha_k$
is obtained by normalizing $\w_k^{(i)}, 1\leq i\leq m$ such that $\Vert
\w_k^{(i)} \Vert$ = 1. With this rank reduction procedure, we can greatly reduce the number of terms in the separated representation of the integrand thus reducing the computation time for quadrature based evaluation of the integrals i.e.
\begin{equation}
I[\tilde{u}_n] \approx I[\tilde{u}_r] = \sum_{k=1}^{r} \alpha_k \left( \prod_{i=1}^{m} \int\limits_{-\infty}^{+\infty} w^{(i)}_k(x_i) \rho^{(i)}(x_i)dx_i\right).
\label{low_rank_integrals}
\end{equation}

\begin{figure}[htb!]
\includegraphics[scale=0.60]{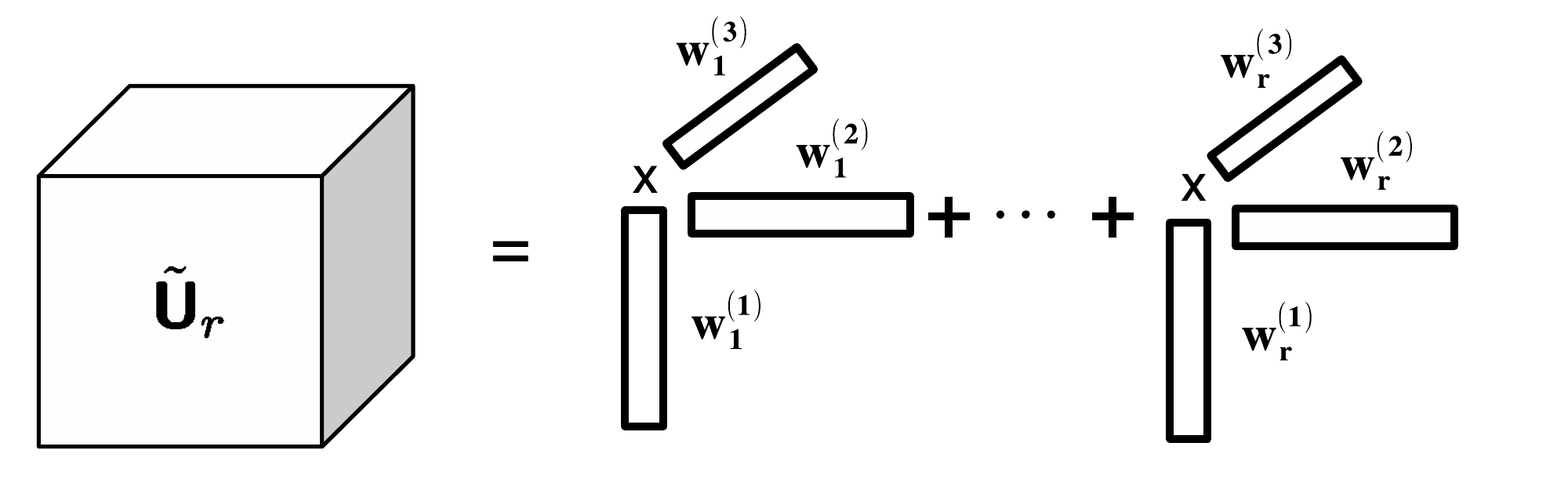}
\caption{Illustration of low rank decomposition of tensor $\tilde{\Ubsf}_r$}
\label{fig:svd3d}       
\end{figure}

In Figure \ref{fig:svd3d}, we illustrate the rank reduction procedure designated
by \eqref{sparse2lr}. Let us denote an $m^{th}$ order tensor $\tilde{\Ubsf}_n$
whose entries are coefficients $v_{\jb}, \;\jb\in \tilde{\Jc}_n$ in the sparse
approximation given by \eqref{lasso}. With choice of bases $\phi_j^{(i)}, 1\leq i\leq m, 1\leq
j\leq p$, function $\tilde{u}_n(\x)$ can be identified with tensor
$\tilde{\Ubsf}_n$. Finding the low rank approximation $\tilde{u}_r(\x)$ in
\eqref{sparse2lr} then reduces to finding a low rank decomposition of the tensor
$\tilde{\Ubsf}_n\approx \tilde{\Ubsf}_r$ such that

\begin{equation}
\tilde{\Ubsf}_r = \sum_{k=1}^{r}\alpha_k\left(\otimes_{i=1}^m \w_k^{(i)}\right).
\label{lowrank}
\end{equation}
The decomposition of the tensor in \eqref{lowrank} is known as the canonical polyadic decomposition for
which several tensor decomposition tools are available. In this work, we use the Tensor Toolbox\cite{TTB_Software}.

We now wish to integrate the low rank function $u_r(\x)$ with respect to a separable measure $\rho(\x)$ via quadrature as
\begin{equation}
I[\tilde{u}_r] = \sum_{k=1}^{r}\alpha_k\left(\prod_{i=1}^m\int_{-\infty}^{+\infty} w^{(i)}_k(x_i)\rho^{(i)} (x_i)dx_i\right),
\label{lowrankquad}
\end{equation}
Let us denote $\gamma_{q_i}^{(i)}$ and $\mathrm{x}_i^{q_i}, 1\leq q_i \leq Q_i$ as one dimensional quadrature weights and quadrature points along dimension $i$ for the measure $\rho^{(i)}(x_i)$. We can evaluate \eqref{lowrankquad} as
\begin{equation}
I[u_r] = \sum_{k=1}^{r}\alpha_k\left(\prod_{i=1}^m\sum_{q_i=1}^{Q_i}\gamma^{(i)}_{q_i} w_k^{(i)}(\mathrm{x}_i^{q_i})\right).
\label{onedquad}
\end{equation}
Thus we integrate our function using $rm$ one dimensional integrals each of which is evaluated using quadrature at a total computational cost of $O(rmp)$, where
$p$ is the degree of univariate polynomial functions $w_k^{(i)}(x_i)$, increasing only linearly with dimension.

\section{Integrals in XVH2}
\label{sec:quantumchem}
In this section, we motivate the utility of sparse low rank tensor based
approximation of PES to find energy corrections and anharmonic frequencies of molecular vibrations.
\subsection{First and second order energy corrections}
\label{sec:XVH2integrals}
Below, we list integrals in the second-order diagrammatic vibrational many-body Green's function formalism. The reader is referred to the original papers
\cite{Hermes:2013,Hermes:2014} for the derivations of this formalism and \cite{RAI17} for a detailed presentation of the following integrals.

In the following, we denote first and second order corrections to energy by $E^{(1)}_0$ and $E^{(2)}_0$
respectively. The anharmonic vibrational frequency of the $i$th mode ($\nu_i$) including up to the second-order perturbation correction can be obtained by frequency-independent, diagonal approximation to the Dyson equation \cite{Hermes:2013}
\begin{eqnarray}
\nu_i = \left\{\omega_i^2 +2\omega_i\Sigma_i(0)\right\}^{1/2},
\label{Dyson_eq1}
\end{eqnarray}
with
\begin{eqnarray}
\Sigma_i (\nu) &=& \Sigma_i^{(1)}(\nu)+ \Sigma_i^{\mathrm{(2p)}} (\nu)
+ \Sigma_i^{\mathrm{(2p^\prime)}} (\nu)+
\nonumber\\
&&+ \; \Sigma_i^{\mathrm{(2b)}} (\nu)
+ \Sigma_i^{\mathrm{(2b^\prime)}} (\nu).
\label{Dyson_eq2}
\end{eqnarray}
where $\omega_i$ is the $i$th harmonic frequency. Note that, in this work, we solve the Dyson equation non-self-consistently, i.e., we substitute $\nu=0$ in the right-hand sides of \eqref{Dyson_eq1} and \eqref{Dyson_eq2}. Given these notations, our interest is in computing the integrals $E_0^{(1)}, E_0^{(2)}, \Sigma_i^{(1)}(0), \Sigma_i^{\mathrm{(2p)}}(0),$  $\Sigma_i^{\mathrm{(2p^\prime)}}(0), \Sigma_i^{\mathrm{(2b)}}(0)$ and  $\Sigma_i^{\mathrm{(2b^\prime)}}(0)$.

We classify all integrals that appear in the formalisms of XVH2 into two groups. The first group
consists of $m$-dimensional integrals ($m=3a-6$, where $a$ is the number of atoms in the non linear molecule), which are collectively written as

\begin{eqnarray}
I^{(1)}= \int_{-\infty}^{+\infty} e(\x) P^{(1)}(\x) \dd\x
\label{I1}
\end{eqnarray}
with
\begin{eqnarray}
e(\x) &=& \prod_{i=1}^m e^{-\omega_ix_i^2}, \label{gaussian} \\
P^{(1)}(\x) &=& \Delta V(\x) \prod_{i=1}^m \lambda_i^{(1)}(x_i), \label{p1}
\end{eqnarray}
where $\x = \{x_1,\dots,x_m\}$ is the $m$-dimensional set of normal coordinates and $\lambda_i^{(1)}(x)$ is given in Table \ref{tab:fo_lambda} for each case of $I^{(1)} = E_0^{(1)}$ or $I^{(1)} = \Sigma_i^{(1)}(0)$.

\begin{table}[h!]
\centering
\caption{The value of $\lambda_i^{(1)}(x_i)$ in Equation (\ref{p1}).}
\begin{tabular}{p{2cm}p{2cm}} \\ \hline
$I^{(1)}$ & $\lambda_{i}^{(1)}(x_i)$\\ \hline
$E_0^{(1)}$ & 1 \\
$\Sigma_i^{(1)}(0)$ & ${2^{1/2}\eta_2(x_i)}/{\eta_0(x_i)}$ \\
$\Sigma_{j \neq i}^{(1)}(0)$ & 1\\
\hline
\end{tabular}
\label{tab:fo_lambda}
\end{table}
Here $\eta_{n_i}(x_i)$ is the harmonic-oscillator wave function along the $i$th normal coordinate $x_i$ with quantum number $n_i$, namely,
\begin{eqnarray}
\eta_{n_i}(x_i) = N_{n_i} e^{-{\omega_i x_i^2}/{2}} h_{n_i}(\omega_i^{1/2}x_i).
\label{eta_fn}
\end{eqnarray}
Here, $N_{n_i}$ is the normalization coefficient, $h_{n_i}$ is the Hermite polynomial of degree $n_i$.

The fluctuation potential $\Delta V(\x)$ is given by
\begin{eqnarray}
\Delta V(\x) = V(\x) -V_{\mathrm{ref}} - \frac{1}{2} \sum_{i=1}^m \omega_i^2 x_i^2,
\end{eqnarray}
where $V(\x)$ is the $m$-dimensional PES and $V_{\mathrm{ref}}$ is its value at
the equilibrium geometry, which is the electronic energy at the equilibrium
geometry of the molecule.

The second group involves $2m$-dimensional integrals of the form,
\begin{eqnarray}
I^{(2)} = \int \limits_{-\infty}^{+\infty}\int \limits_{-\infty}^{+\infty} e(\x,\x') P^{(2)}(\x,\x')
\dd\x \dd\x' \label{reform_E2}
\end{eqnarray}
with
\begin{eqnarray}
e(\x,\x') &=& \prod_{i=1}^m e^{-\omega_i(x_i^2+x'^{2}_i)}, \label{gaussian2} \\
P^{(2)}(\x,\x') &=& \Delta V(\x) \Delta V(\x') G(\x,\x')\prod_{i=1}^m \lambda_i^{(2)}(x_i,x_i'), \label{p2}
\end{eqnarray}
where $G(\x,\x')$ is a real-space Green's function given by
\begin{eqnarray}
&& G(\x,\x') = \underset{(n_1,n_2,\dots,n_m) \neq (0,0,\dots,0)}{\sum_{n_1=0}^{n_{\mathrm{max}}}\cdots\sum_{n_m=0}^{n_{\mathrm{max}}}}\prod_{i=1}^m \frac{N^2_{n_i} h_{n_i}(\omega_i^{1/2}x_i)h_{n_i}(\omega_i^{1/2}x'_i)}
{-\sum_{i=1}^{m} n_i \omega_i} . \label{generalG}
\end{eqnarray}

Table \ref{tab:so_lambda} defines $\lambda_i^{(2)}(x_i,x'_i)$ for each case under consideration.
\begin{table}[h!]
\centering
\caption{The value of $\lambda_i^{(2)}(x_i,x'_i)$ in Equation (\ref{generalG}).}
\begin{tabular}{ ll } \\ \hline
$I^{(2)}$ & $\lambda_i^{(2)}(x_i,x'_i)$\\
\hline
$E_0^{(2)}$ & 1 \\
$\Sigma_i^{\mathrm{(2p)}}(0)$ & ${(n_i+2)^{1/2}(n_i+1)^{1/2}\eta_{n_i+2}(x_i')}/{\eta_{n_i}(x_i')}$\\
$\Sigma_{j \neq i}^{\mathrm{(2p)}}(0)$ & 1\\
$\Sigma_i^{\mathrm{(2b)}}(0)$ & ${(n_i+1)\eta_{n_i+1}(x_i)\eta_{n_i+1}(x_i')}/\{{\eta_{n_i}(x_i)\eta_{n_i}(x_i')\}}$\\
$\Sigma_{j \neq i}^{\mathrm{(2b)}}(0)$ & 1\\ \hline
\end{tabular}
\label{tab:so_lambda}
\end{table}

Thus, the integrand factor $P^{(2)}$ in~\eqref{p2} is a polynomial in $\x$ and $\x'$. It is clear that the dimensions
of integrands in \eqref{I1} and \eqref{reform_E2} grow linearly with the number of
atoms in a molecule. With this integration problem at hand, we briefly explain our
approach of separated integration for both $I^{(1)}$ and $I^{(2)}$.

In $I^{(1)}$, the weight function $e(\x)$ is a separable Gaussian, and $P^{(1)}(\x)$ has the factor of $\Delta V(\x)$ which can be expressed in a low-rank format
\begin{eqnarray}
\Delta V(\x) \approx \sum_{k=1}^{r_1} \prod_{i=1}^m \Delta V^{(i)}_k(x_i)
\label{lr_dv}
\end{eqnarray}
with a separation rank $r_1$.
Thus $I^{(1)}$ can be evaluated as a sum-of-products of one-dimensional integrals,
\begin{eqnarray}
I^{(1)} \approx \sum_{k=1}^{r_1} \prod_{i=1}^m \int \limits_{-\infty}^{+\infty} e^{-\omega_i
x_i^2}\Delta V^{(i)}_k(x_i) \lambda_{i}^{(1)}(x_i)\,\dd x_i,
\label{sep_E1}
\end{eqnarray}
using Gauss-Hermite quadrature at a cost $O(r_1mp)$ that scales linearly with the dimension $m$.

In $I^{(2)}$, $G(\x,\x')$ appears as an integrand factor which can also be
represented with a low rank approximation
\begin{eqnarray}
G(\x,\x') \approx \sum_{k=1}^{r_2} \prod_{i=1}^m G^{(i)}_k(x_i,x'_i),
\label{lowrankG}
\end{eqnarray}
with a separation rank $r_2$ (see \cite{RAI17}, section IV-D for a detailed discussion).
Substituting \eqref{lr_dv} and \eqref{lowrankG} in \eqref{reform_E2}, $I^{(2)}$ can be evaluated as a sum-of-products of two-dimensional integrals,
\begin{equation}
I^{(2)} \approx \sum_{k_1=1}^{r_1} \sum_{k_2=1}^{r_1} \sum_{k_3=1}^{r_2} \prod_{i=1}^m
\int_{-\infty}^{+\infty} \int_{-\infty}^{+\infty} e^{-\omega_i(x_i^2+x_i'^2)}
\Delta V^{(i)}_{k_1}(x_i) \Delta V^{(i)}_{k_2}(x'_i) \, G_{k_3}^{(i)}(x_i,x'_i) \lambda_i^{(2)}(x_i,x_i')\,\dd x_i \,\dd x_i',
\label{sep_E2}
\end{equation}
at a cost $O(r_1^2r_2mp)$ that again scales linearly with dimension, albeit with a larger prefactor compared to the cost of $I^{(1)}$.

For accurate, more efficient and scalable computation of $I^{(1)}$ and $I^{(2)}$ using separated integration with \eqref{sep_E1} and \eqref{sep_E2}, we require two conditions to be satisfied. Firstly, the low rank approximations in \eqref{lr_dv} should be sufficiently accurate using as few samples of $\Delta V(\x)$ as possible. We achieve this objective by sparse approximation using compressed sensing explained in Section \ref{sec:sparse}. Secondly, the separation rank $r_1$ in \eqref{lr_dv} and $r_2$ in \eqref{lowrankG} must be small for sufficiently accurate approximation in order to reduce computation time of quadrature integration of the associated one- or two-dimensional integrals. We achieve this by low rank compression explained in Section \ref{sec:low_rank}.

We call the developed XVH2 method that uses the sparse low-rank-decomposed PES and Green's function presented in this study the sparse canonical-tensor XVH2 (SCT-XVH2) method. The overall algorithm is outlined briefly in Table \ref{alg:sct-xvh2}.

\begin{table}[htbp!]
\caption{An outline of SCT-XVH2.}
\label{alg:sct-xvh2}
\begin{tabular}{l}
\hline
\textbf{Input:} $r_1$, $r_2$, $S$, $p$, $\Delta V(\mathbf{x}^s)$, $\phi^{(i)}_{j_i}(\mathrm{x}_i^{s_i})$\\
\textbf{Output:} $E^{(1)}_0, E^{(2)}_0,\nu_i$\\
\textbf{1:} Get sparse approximation of $\Delta V(\x)$ by a compressed sensing software as explained in Section \ref{sec:sparse}.\\
\textbf{2:} Get low rank representation of sparse $\Delta V(\x)$ \eqref{lr_dv} by using a tensor decomposition\\ \;\;\;\;  software as explained in Section \ref{sec:low_rank}.\\
\textbf{3:} Obtain low rank compression of Green's function \eqref{lowrankG} using the same tensor decomposition \\ \;\;\;\;\; software.\\
\textbf{4:} Calculate $E_0^{(1)}, \Sigma_i^{(1)}(0)$ by solving \eqref{sep_E1} using appropriate $\lambda^{(1)}_i(x_i)$ from Table \ref{tab:fo_lambda}. \\
\textbf{5:} Calculate $E_0^{(2)}, \Sigma_i^{\mathrm{(2p)}}(0),  \Sigma_i^{\mathrm{(2p^\prime)}}(0), \Sigma_i^{\mathrm{(2b)}}(0)$ and  $\Sigma_i^{\mathrm{(2b^\prime)}}(0)$ by solving \eqref{sep_E2} using appropriate \\ \;\;\;\;  $\lambda^{(2)}_i(x_i,x_i')$ from Table \ref{tab:so_lambda}. \\
\textbf{6:} Calculate $\nu_i$ using Dyson equations \eqref{Dyson_eq1} and  \eqref{Dyson_eq2}.\\
\hline
\end{tabular}
\end{table}

\section{Results}
\label{sec:applications}

In this section, we illustrate the application of the above method for
approximation of PES of molecules and integrals for estimation of zero point
energies and frequencies. In the first subsection, we discuss sparse
approximation of the PES of water, formaldehyde, methane and ethylene using
compressed sensing. Next, we present results on low rank compression of sparse
PES and Green's function. Finally, we apply the method of separated integrals in XVH2 for
estimating anharmonic zero point energies and frequencies. We compare this method with
other methods including our previous method CT-XVH2 \cite{RAI17}.

\subsection{PES approximation using Compressed Sensing}
To estimate error in approximation of the PES, we consider a separate test sample
set with $N_{Test}$ evaluations of the PES to determine the accuracy of the sparse
approximation. Similar to \cite{HARDING17}, both training and test sets were sampled uniformly with energies less than $45$ kcal/mol above the global minimum. The relative approximation error of the sparse solution is defined as
\begin{equation}
\epsilon_s = \frac{\Vert \u - \tilde{\u}_n\Vert_2}{\Vert \u \Vert_2},
\label{epsilon_s}
\end{equation}
where $\u$ is a vector containing function evaluations of the test set and $\tilde{\u}_n$ is a vector
that contains corresponding evaluations of $\tilde{u}_n(\x)$. We also
compare it with our previous approach, named CT-XVH2, of direct approximation (i.e. without prior step of sparse approximation) of PES in canonical low rank tensor format~\cite{RAI17}.
We consider four molecules, water (H$_2$O, $m=3$), formaldehyde (H$_2$CO, $m=6$), methane (CH$_4$, $m=9$) and ethylene (C$_2$H$_4$, $m=12$). The potential energy evaluations are obtained from the software package NWChem \cite{nwchem} at MP2/aug-cc-PVTZ electronic structure theory. For the approximation basis, we choose orthogonal Hermite polynomial basis functions with total degree $p=6$, i.e. the set of interest is $\mathbb{P}_6$. In order to take variation due to sampling effects into account, we illustrate $\epsilon_s$ in plotted results by connecting the median relative errors while each error bar is indicating the 25/75 quantiles obtained from 51 independent sample sets.

\begin{figure}[htb!]
\includegraphics[scale=.35]{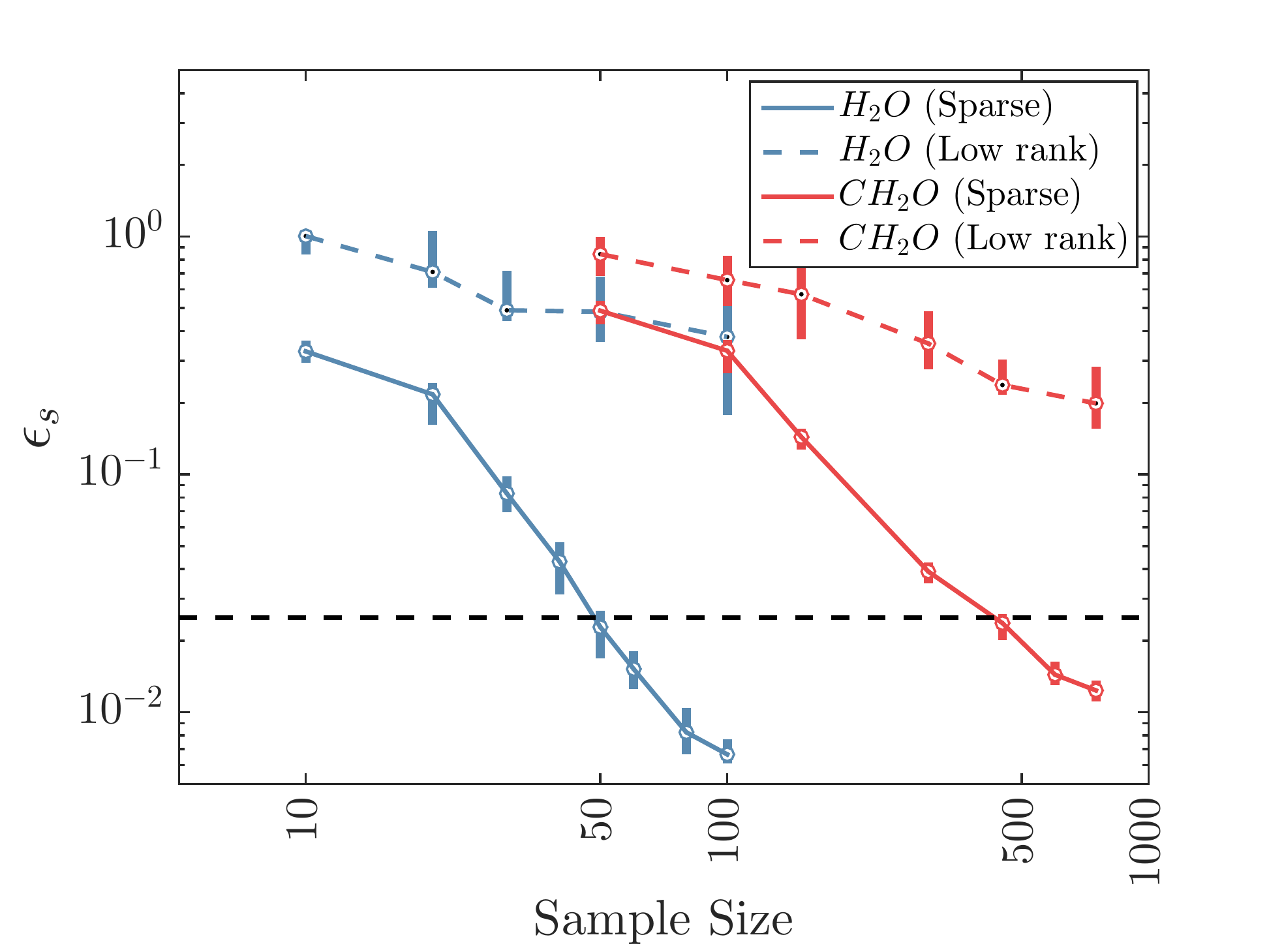}
\includegraphics[scale=.35]{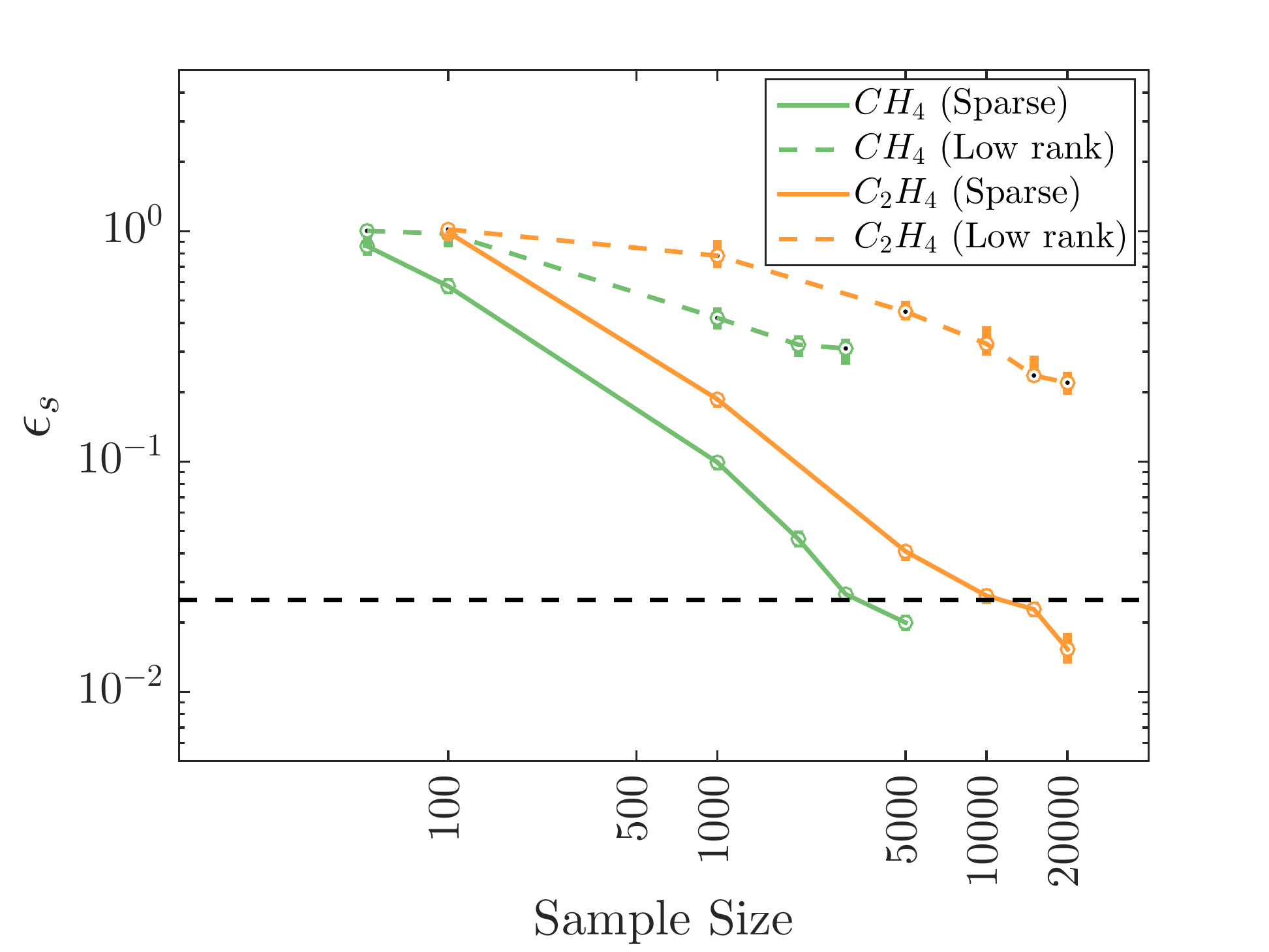}
\caption{The relative approximation error $\epsilon_s$ in PES
of (a) water and formaldehyde and (b) methane and ethylene as a function of the sample size using compressed sensing (SCT-XVH2) and canonical low rank tensor (CT-XVH2) methods. The plot connects the medians, with each error bar indicating the 25/75 quantiles obtained from 51 independent ensembles. Black dotted lines determine the number of samples for an accuracy of 2.5\%.}
\label{fig:sparse_pes}       
\end{figure}

\begin{figure}[htb!]
\includegraphics[scale=.34]{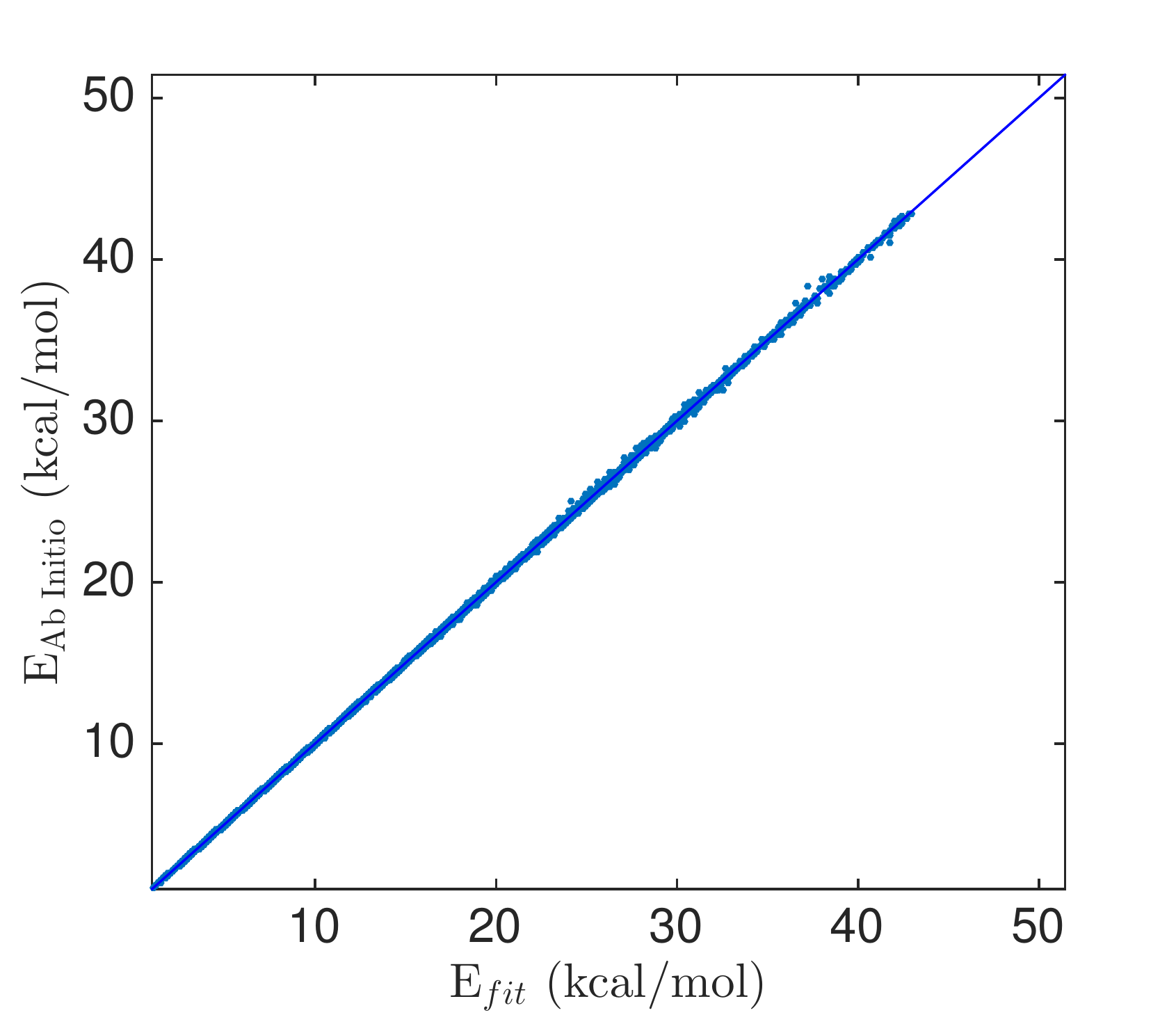}
\includegraphics[scale=.34]{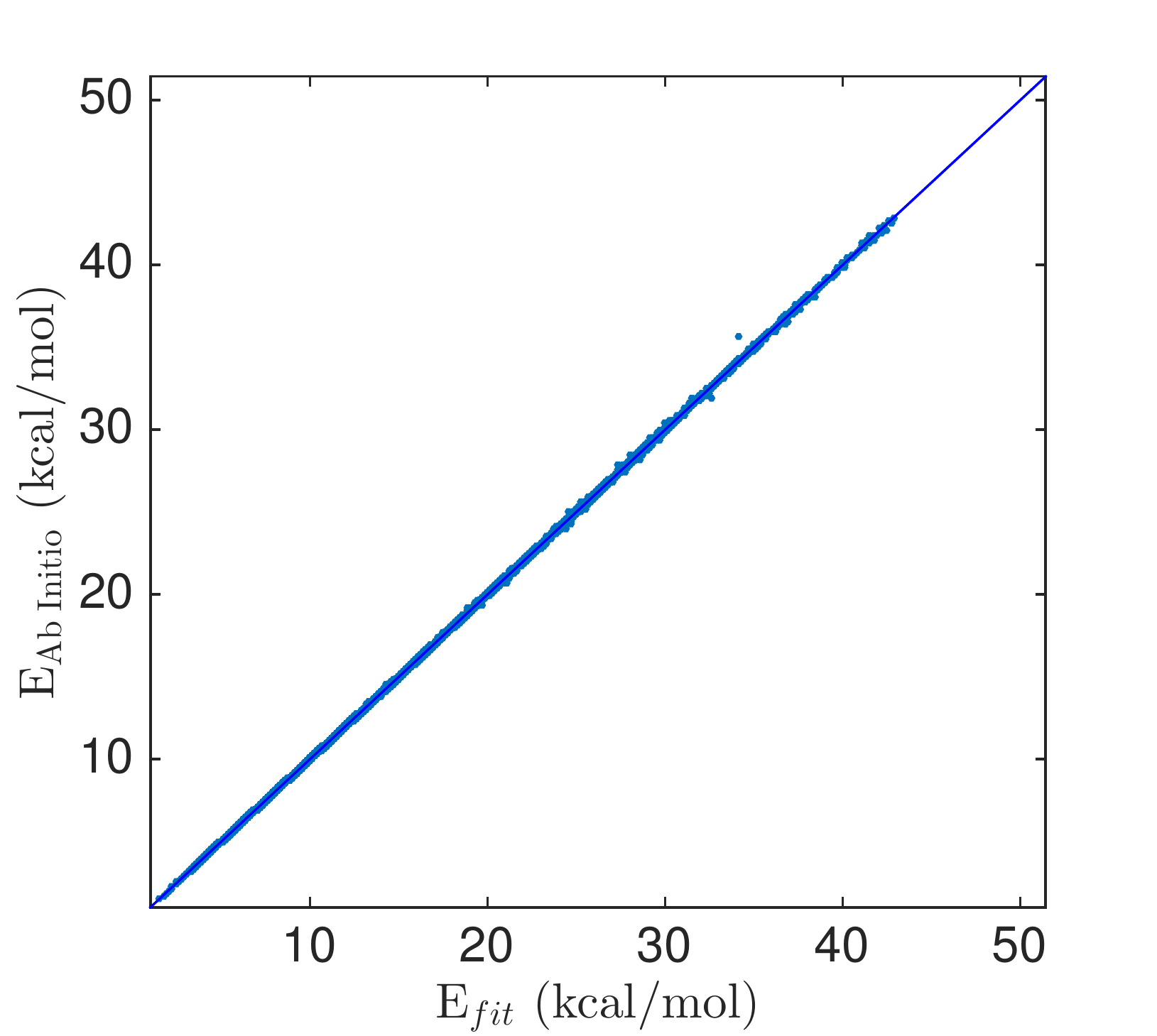}
\caption{Scatter plot of the ab initio energies vs fitted energies using compressed sensing of 10000 test points for methane potential surface using $S = 3000$ and ethylene potential surface using $S = 12000$ points.}
\label{fig:sparse_scatter}       
\end{figure}

Figure \ref{fig:sparse_pes} plots the relative error of the sparse approximation
of the PES, $\epsilon_s$ [Equation \eqref{epsilon_s}], as a function of sample
size $S$. For the sake of comparison, we also show error obtained using direct
approximation in canonical low rank tensor format of rank $\leq 30$ (as proposed
in CT-XVH2 \cite{RAI17}). We first note that, with increase in sample size $S$,
$\epsilon_s$ is reduced dramatically for all the molecules. In contrast, PES
approximated as canonical low rank tensor has a slower convergence rate for the
same sample sizes. In order to determine the desired accuracy of the sparse PES, 
we determine the number of samples required to fit the surface with the same accuracy 
as reported in \cite{HARDING17}. We take both methane and ethylene as typical examples for estimating the accuracy of fit. For both molecules, we sampled the surface with $N_{Test} = 10000$ points that cover the range up to 15000 cm$^{-1}$ above the minimum. Figure \ref{fig:sparse_scatter} shows scatter plots of methane and ethylene comparing ab initio and fitted potential energy surfaces for the test points. The points within 1000 cm$^{-1}$ of minimum are fit with a RMS error of 0.67 cm$^{-1}$ and 1.2 cm$^{-1}$ for methane and ethylene respectively. The RMS error for all 10000 test points is 18.5 cm$^{-1}$ and 12.5 cm$^{-1}$ respectively.
This suggests that a relative accuracy of $\approx 2.5\%$ in
approximation of PES is sufficient for accurate estimates of the quantities of
interest (zero point energy and frequencies corrections). In the case of water and
formaldehyde (Figure \ref{fig:sparse_pes}(a)), to achieve the same accuracy, we need
$S = 50$ and $S = 450$  respectively using compressed sensing, at least an order
of magnitude smaller than direct approximation in canonical low rank tensor
format. Similarly, as shown in Figure \ref{fig:sparse_pes}(b),  $S = 3000$ for methane and $S = 12000$ for
ethylene while the PES expressed as canonical low rank tensor is an order of
magnitude less accurate.

Table \ref{tab:pes_approx} summarizes our results related to sparse
approximation of the PES of molecules using compressed sensing. Note that
choosing the basis set $\mathbb{P}_6$ (instead of $\mathbb{Q}_6$) results in
drastic reduction in the number of basis functions for approximation of the PES
using least-squares with $\ell_1$ regularization in compressed sensing. The reduction is by
several orders of magnitude for high dimensional PES (as in methane and ethylene
molecules) and choosing the basis set $\mathbb{P}_p$ could very well be
indispensable for higher values of $m$. Note that, for all four molecules, the
number of samples ($S$) is smaller than number of basis functions in
$\mathbb{P}_6$ (compare values in row 3 and 5 in Table \ref{tab:pes_approx}).
Intuitively, such a reduction is achieved because $\ell_1$ regularization in
compressed sensing in \eqref{sparse_one} effectively exploits sparsity
structure, thereby allowing more accurate estimation of the model parameters. Indeed such an approach
will not result in a good approximation of the function if the latter is not
sparse on the chosen polynomial basis.

\begin{table}[htb!]
\caption{Summary of sparse approximation of PES for four molecules with increasing dimensionality of PES}
\label{tab:water}
\begin{tabular}{p{3.5cm}p{1cm}p{1.8cm}p{1.8cm}p{1.8cm}}
\hline\noalign{\smallskip}
 & $\mathrm{H_2O}$& $\mathrm{CH_2O}$ & $\mathrm{CH_4}$ & $\mathrm{C_2H_4}$\\
\noalign{\smallskip}\hline\noalign{\smallskip}
$m$ & 3 & 6 & 9 & 12\\
Number of basis in $\mathbb{Q}_{6}$ & 343 & $1.17\times 10^{5}$ & $4.03 \times 10^{7}$ & $1.38 \times 10^{10}$\\
Number of basis in $\mathbb{P}_{6}$ & 84 & 924 & 5005 & 18564\\
$N_{Test}$ & 100 & 1000 & 10000 & 10000 \\
$S$ & 50 & 450 & 3000 & 12000 \\
\noalign{\smallskip}\hline\noalign{\smallskip}
\end{tabular}
\label{tab:pes_approx}
\end{table}

\subsection{Low rank compression of sparse PES and Green's function}
\begin{figure}[htb!]
\includegraphics[scale=.35]{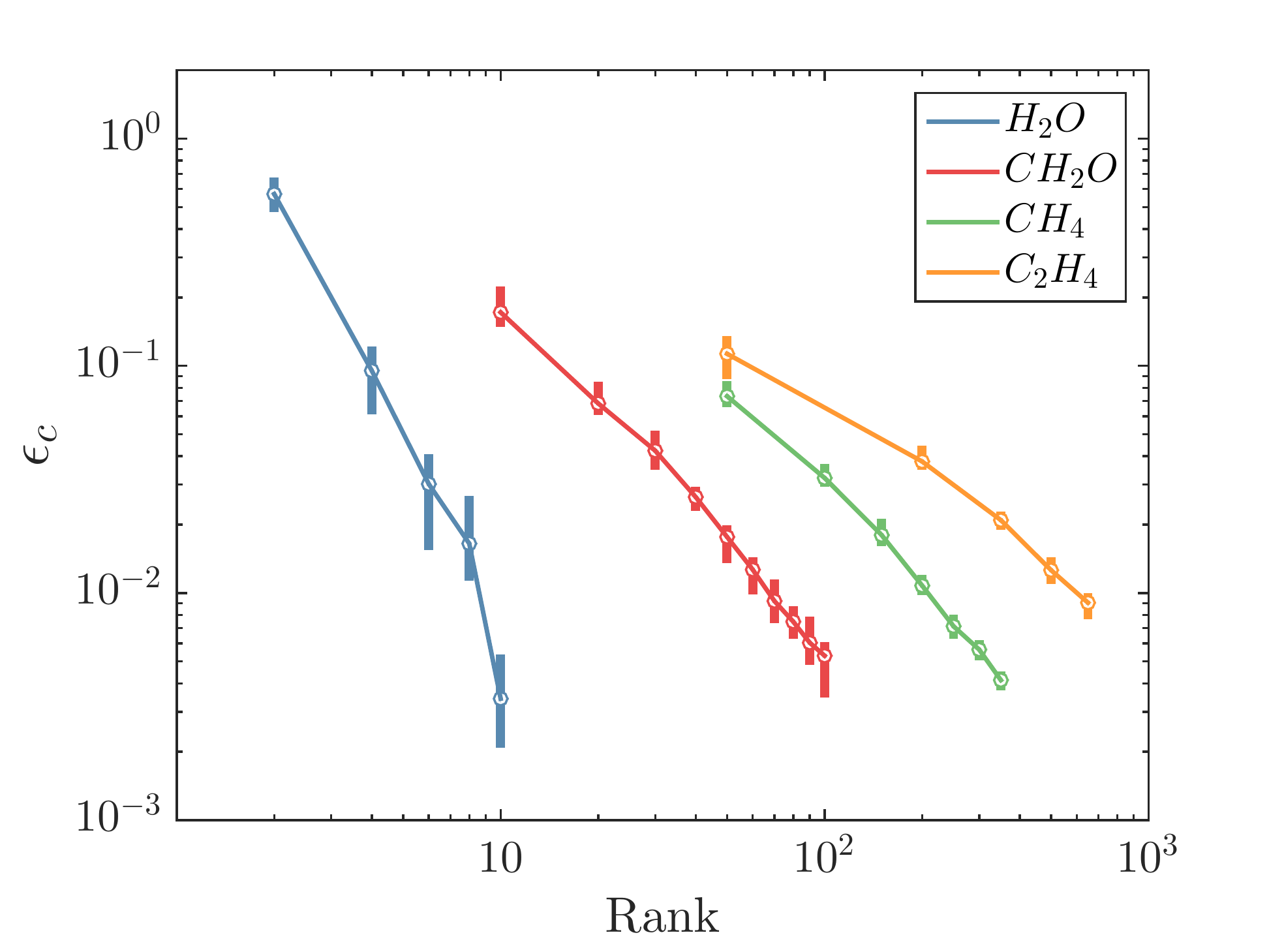}
\includegraphics[scale=.35]{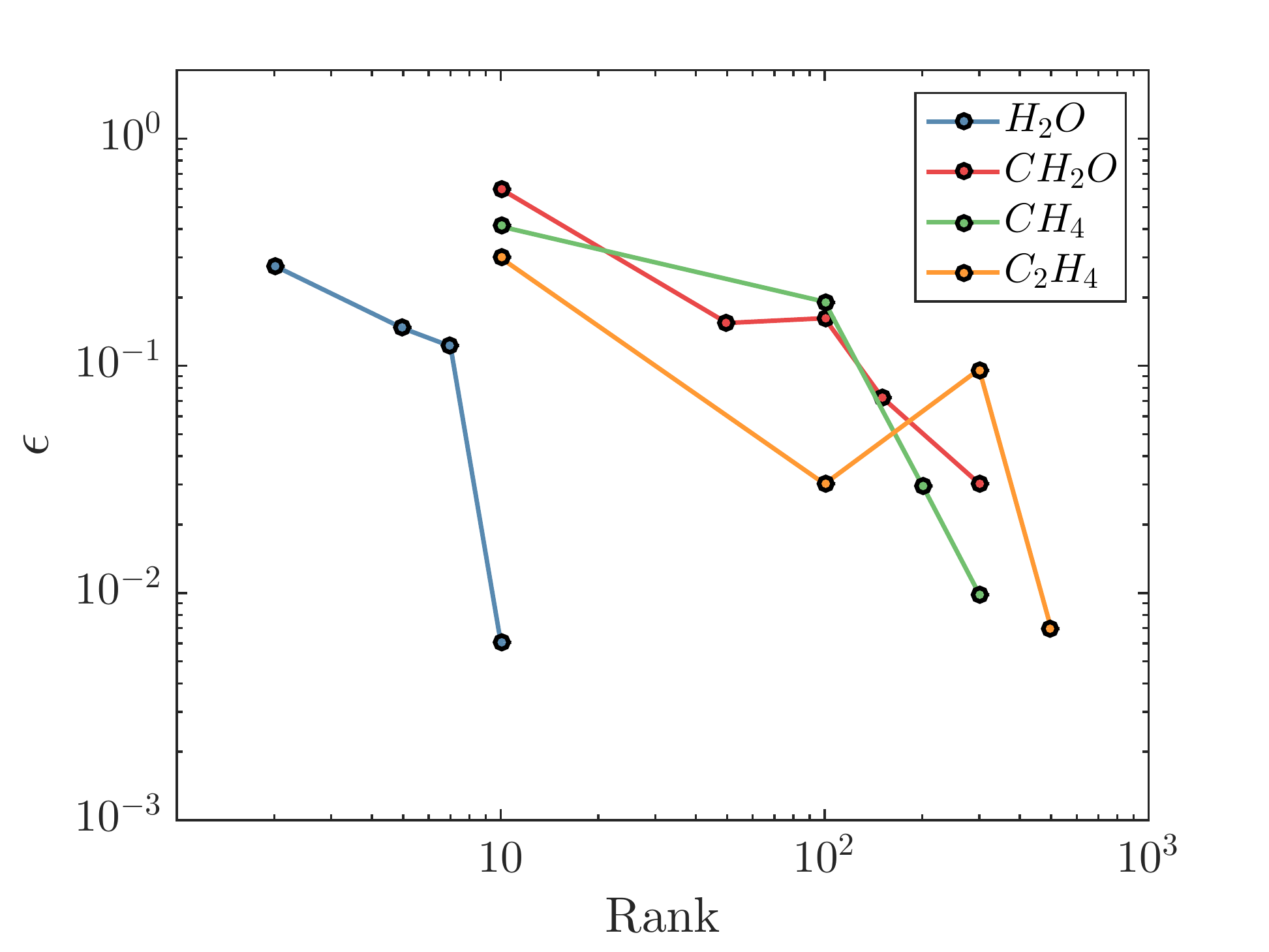}
\caption{(a) The compression error $\epsilon_c$ in low rank PES
of molecules as a function of separation rank. The plot connects the medians, with each error bar indicating the 25/75 quantiles obtained from 51 independent ensembles. The error bars correspond to compression of 51 sparse tensors obtained from the sparse approximation algorithm. (b) Compression error of Green's function as a function of separation rank.}
\label{fig:compress}
\end{figure}

In this section, we discuss results related to low rank compression of integrands in XVH2: sparse PES obtained using compressed sensing and Green's function, one of the integrand factors in XVH2 integrals. We define the error due to low rank compression as
\begin{equation}
\epsilon_c = \frac{\Vert \tilde{\u}_n - \tilde{\u}_r\Vert_2}{\Vert \tilde{\u}_n \Vert_2},
\label{epsilon_c}
\end{equation}
where $\tilde{\u}_r$ is a vector containing evaluations of $\tilde{u}_r(\x)$ as
defined in Eq.~\eqref{sparse2lr} at test sample points. Thus the value of
$\epsilon_c$ indicates how accurately the low rank decomposed PES
approximates the sparse PES. Note that we use the same test samples in
\eqref{epsilon_s} and \eqref{epsilon_c}. As in previous plot, we illustrate
$\epsilon_c$ in plotted results by connecting the median error with each error
bar indicating the 25/75 quantiles obtained from 51 independent sample sets.
Note that we have error bars on $\epsilon_c$ because the sparse approximation
algorithm will estimate a different set of coefficients for each independent
sample set of a given size.  We show the application of low rank compression of
the sparse PES in Figure \ref{fig:compress} (a), where we plot the compression
error versus separation rank for water, formaldehyde, methane and ethylene
molecules. Table \ref{tab:compression_results} shows the numbers pertaining to
low rank compression of PES and Green's function for separated integration.

In Figure \ref{fig:compress} (a), we find that the compression error reduces
with increasing separation rank for all four molecules. In Table
\ref{tab:compression_results}, we summarize the ranks of PES before and after
compression (rows one and two respectively). Note that the separation rank of
PES before compression is the same as number of non-zero coefficients in the
sparse PES. For rank corresponding to
$\epsilon_c<10^{-2}$ in Figure \ref{fig:compress}(a), the proposed low rank
compression is able to achieve approximately a reduction in the separation rank
by a factor of three, for a small loss of accuracy. Depending on the available
computational budget and required accuracy, this procedure provides flexibility
in truncating the separation rank of the PES, thus providing more flexibility in
estimating zero-point energy and frequency corrections.

\begin{table}[hbt!]
\caption{Summary of low rank compression of sparse PES and Green's function}
\label{tab:water}       
%
%
\begin{tabular}{p{4cm}p{1cm}p{1.4cm}p{1.4cm}p{1.4cm}}
\hline\noalign{\smallskip}
 & $\mathrm{H_2O}$& $\mathrm{CH_2O}$ & $\mathrm{CH_4}$ & $\mathrm{C_2H_4}$\\
\noalign{\smallskip}\hline\noalign{\smallskip}
PES rank before compression & 39 & 244 & 996 & 1702 \\
PES rank after compression for $\epsilon_c<10^{-2}$ & 10 & 100 & 300 & 500 \\
Quantum number $n_{max}$ in Green's function & 9 & 7 & 5 & 4\\
Rank of compressed Green's function & 10 & 300 & 300 & 500 \\
Green's function compression ratio $(\gamma)$ &0.37 & 0.10 & $6.9\times10^{-3}$ & $1.4\times10^{-3}$\\
\noalign{\smallskip}\hline\noalign{\smallskip}
\end{tabular}
\label{tab:compression_results}
\end{table}

In \cite{RAI17} (see section IV.D), we introduced low rank compression of the
Green's function in CT-XVH2 for estimating zero-point energies and frequencies
of water and formaldehyde molecules. Here we extend it to methane and ethylene
albeit with smaller values of maximum quantum number $n_{max}$. Unlike the sparse
PES, the tensor representation of Green's function is full and requires storage
of $n_{max}^m$ coefficients thus constraining the maximum value of $n_{max}$
(see row 3 of Table \ref{tab:compression_results} for values of $n_{max}$
considered for each molecule). Figure \ref{fig:compress}(b) shows the
compression error of $G(\x,\x')$ in \eqref{generalG} as a function of the
separation rank $r$ for each of the four molecules. We
find that the low-rank compression of $G(\x,\x')$ for each molecule is obtained
with very small separation ranks as compared to full multidimensional functional representation
in \eqref{generalG} (see row 4 of Table
\ref{tab:compression_results} for exact values of separation ranks). Note that
one can choose higher compression rank to further reduce $\epsilon$. Also, the convergence of $\epsilon$ with rank could be non-monotonic (e.g. for ethylene (C$_2$H$_4$) in Figure \ref{fig:compress}(b)) since the alternating minimization scheme employed for compression is an approximate algorithm that is not guaranteed to find optimal approximation for a given rank. We
measure the compression efficiency using compression ratio defined as the ratio
of the number of parameters in the low-rank-decomposed (i.e., compressed)
representation of $G(\x,\x')$ to the total number of parameters in the original
representation:
\begin{equation}
 \gamma = \frac{r n_{max} m}{n_{max}^m}
\end{equation}
The values of $\gamma$ for each molecule is reported in Table \ref{tab:compression_results} (see last row). We find that the compression ratio decreases progressively from smaller to bigger molecule indicating that $G(\x,\x')$ is greatly compressed without any significant loss of information.

Having obtained accurate compressed representations of the sparse PES and Green's function, we report zero-point energies and frequencies of each molecule in the following subsection.

\subsection{Anharmonic zero-point energies and frequencies of molecules}

In this section, we compare the cost and accuracy of SCT-XVH2 with those of
CT-XVH2 \cite{RAI17}, Monte Carlo XVH2 (MC-XVH2) and XVH2 for the first- and
second-order corrections to the zero-point energies and frequencies of water and
formaldehyde \cite{Hermes:2013, Hermes:2015}. For methane and ethylene, we
simply report the results obtained using SCT-XVH2. In comparing these methods,
we compare two scenarios depending on the method of PES evaluations: direct and
indirect. In indirect evaluations, the PES is given as a quartic force field.
Accordingly, the methods using a QFF in the following are abbreviated with a
parenthesis with the truncation rank of the Taylor-series PES, which is 4 in
this case. The XVH2 results are considered to be benchmark for methods using
indirect PES evaluations, having only roundoff errors. In direct calculations,
the value of PES at a given geometry is calculated on demand by NWChem at the
MP2/aug-cc-pVTZ electronic structure theory. All calculations solve the Dyson
equation non-self-consistently.

\begin{table}[htb!]
\caption{The first- and second-order anharmonic corrections to the zero-point energy [$E_0^{(1)}$ and $E_0^{(2)}$] and frequencies of the three fundamental transitions $(\nu_i)$ in $\mathrm{cm}^{-1}$ of the water molecule. The separation rank of PES in SCT-XVH2 and Green's function is 10.}
\label{tab:water}       
%
%
\begin{tabular}{p{0.7cm}p{2cm}p{2.2cm}p{2cm}p{2cm}p{2cm}}
\hline\noalign{\smallskip}
 & SCT-XVH2 & SCT-XVH2(4) & CT-XVH2(4) & MC-XVH2(4) & XVH2(4)\\
\noalign{\smallskip}\hline\noalign{\smallskip}
$S$ & 50 & 35 & 150 & $7 \times 10^{5}$ & 1296 \\
 $E_0^{(1)}$& $52.6 \pm 0.9$& $51.6\pm 0.01$& $51.5\pm0.3$ & $51.3\pm1.1$ & 51.6\\
 $E_0^{(2)}$& $-124.4 \pm 0.9$& $-120.6 \pm 0.0$& $-120.5 \pm 0.2$ & $-119.1\pm0.7$ & $-120.6$ \\
 $\nu_1$& $3616.7 \pm 7.2$&$3644.00 \pm 0.05$&$3645.3 \pm 0.6$ & $3646.9\pm4.5$ & $3645.1$ \\
 $\nu_2$&$1573.7 \pm 1.4$&$1566.88 \pm 0.02$&$1566.7 \pm 0.5$ & $1566.3\pm2.1$ & $1566.9$ \\
 $\nu_3$&$ 3736.2 \pm 3.0$&$3767.28 \pm 0.04$&$3767.3 \pm 0.5$ & $3768.5\pm3.2$ & $3767.4$ \\
\noalign{\smallskip}\hline\noalign{\smallskip}
\end{tabular}
\end{table}

Let us first compare methods with indirect PES evaluations for the water
molecule in Table \ref{tab:water}. Using SCT-XVH2(4), with only 35 samples, the
zero point energies and fundamental frequencies are accurate to within 1
$\mathrm{cm}^{-1}$ of the correct XVH2(4) results. Also, based on the number of
required PES evaluations, SCT-XVH2(4) is four times as fast as CT-XVH2(4)
\cite{RAI17}, two orders of magnitude more efficient than XVH2(4) and three
orders of magnitude less costly than MC-XVH2(4) \cite{Hermes:2015}. In direct
PES evaluations, with only 50 PES samples, SCT-XVH2 calculates the zero
point energies and frequencies while taking into account possibly
higher-than-quartic force constants as compared to other indirect methods that
use QFF. It is twice as costly as SCT-XVH2(4) which uses a QFF. The difference
between the results of SCT-XVH2 and other indirect methods can be attributed to
the fact that the maximum degree of polynomial basis for approximation of PES
using compressed sensing is not restricted to 4 as in QFF.

\begin{table}
\caption{The first- and second-order anharmonic corrections [$E_0^{(1)}$ and $E_0^{(2)}$] to the zero-point energy and the anharmonic frequencies of fundamental transitions $(\nu_i)$ in $\mathrm{cm}^{-1}$ of the formaldehyde molecule. The separation rank of PES in SCT-XVH2 is 100 and Green's function is 300.}
\label{tab:formaldehyde}       
%
%
\begin{tabular}{p{1cm}p{2cm}p{2.4cm}p{2cm}p{2cm}p{2cm}}
\hline\noalign{\smallskip}
 & SCT-XVH2& SCT-XVH2(4)&CT-XVH2(4) & XVH2(4)\\
\noalign{\smallskip}\hline\noalign{\smallskip}
$S$ & 450 & 180 & 4000 &  20736 \\
 $E_0^{(1)}$& $-0.5 \pm 0.5$ &$-0.9 \pm 0.5$&$-1.1 \pm 0.3$ & $-1.0$  \\
 $E_0^{(2)}$& $-77.6 \pm 0.3$& $-73.8 \pm 0.5$& $-77.7 \pm 0.3$ & $-77.7$ \\
 $\nu_1$& $2801.8 \pm 1.8$ &$2810.7 \pm 1.8$ &$2810.6 \pm 0.5$ & $2810.7$ \\
 $\nu_2$& $1723.5 \pm 1.2$ & $1722.2\pm 1.8$& $1723.2\pm 0.3$ & $1723.3$ \\
 $\nu_3$& $1510.4 \pm 1.6$ &$1507.8 \pm 2.1$&$1506.4 \pm 0.7$ & $1506.3$ \\
 $\nu_4$& $1171.0 \pm 1.3$ & $1169.2 \pm 1.2$ & $1166.4 \pm 0.4$ & $1166.4$\\
 $\nu_5$& $2864.1 \pm 2.9$ &$2870.3 \pm 1.9$ &$2870.4 \pm 0.9$ & $2870.8$ \\
 $\nu_6$& $1246.3 \pm 1.2$ &$1244.1 \pm 1.8$&$1243.1 \pm 0.8$ & $1243.0$\\
\noalign{\smallskip}\hline\noalign{\smallskip}
\end{tabular}
\end{table}

In comparing indirect methods for formaldehyde (Table \ref{tab:formaldehyde}),
with as low as 180 PES evaluations in SCT-XVH2(4), the zero-point energies and
frequencies are within 1 $\mathrm{cm}^{-1}$ of the XVH2(4) results, which
require more than 20000 samples. Also, CT-XVH2(4) needs 4000 evaluations to get
comparable estimates. Therefore, SCT-XVH2(4) is an order of magnitude more
efficient than CT-XVH2(4) and two orders of magnitude less costly than
XVH2(4) to reach practically useful accuracy of 1 $\mathrm{cm}^{-1}$.
Using direct PES evaluations, SCT-XVH2 requires 450 samples to estimate the same
quantities while possibly taking into account higher order terms in the
representation of PES as compared to QFF.

In Table \ref{tab:methane_ethylene}, we report the zero-point energies and
frequencies of methane and ethylene. We compare zero point energy corrections of these molecules with 
ones reported in \cite{HARDING17} where Diffusion Monte Carlo (DMC) technique is introduced for estimating 
anharmonic corrections to energy. For methane, the resultant anharmonic correction to zero point energy using SCT-XVH2 ($E_0^{(1)}+E_0^{(2)} $) is $-150.9 \pm 0.8$ cm$^{-1}$ as compared to -143 cm$^{-1}$ in DMC. Similarly, for ethylene, we estimate anharmonic correction to be $-151.6\pm2$ cm$^{-1}$ using SCT-XVH2 as compared to -143 cm$^{-1}$ with DMC. The deviation in corrections, less than 10 cm$^{-1}$ for both molecules, using the two methods is attributed to two factors. Firstly, the electronic structure theory in SCT-XVH2 for PES evaluations is MP2/aug-cc-pVTZ whereas energy corrections reported in \cite{HARDING17} is obtained with CCSD(T)/cc-pVTZ. Secondly, due to limitations on storage and low rank compression of Green's function, we constrained the values of maximum quantum number $n_{max}=5$ for methane and $n_{max}=4$ for ethylene. A method to overcome this limitation is briefly discussed in section \ref{sec:conclusion}.
\begin{table}
\centering
\caption{The first- and second-order anharmonic corrections [$E_0^{(1)}$ and $E_0^{(2)}$] to the zero-point energy and the anharmonic frequencies of fundamental transitions $(\nu_i)$ in $\mathrm{cm}^{-1}$ of methane and ethylene molecule using SCT-XVH2. The separation rank of PES and Green's function for methane is 300 and ethylene is 500.}
\label{tab:methane_ethylene}       
%
%
\begin{tabular}{p{2cm}p{2.4cm}p{2.4cm}}
\hline\noalign{\smallskip}
 & Methane $(\mathrm{CH}_4)$& Ethylene $(\mathrm{C_2H_4})$ \\
\noalign{\smallskip}\hline\noalign{\smallskip}
$S$ & 3000 & 12000 \\
 $E_0^{(1)}$& $-14.6 \pm 0.6$ & $-40.6 \pm 1.7$\\
 $E_0^{(2)}$& $-136.3 \pm 0.2$& $-111.0 \pm 0.3$ \\
 $\nu_1$& $1311.4 \pm 0.6$ & $822.8 \pm 1.0$ \\
 $\nu_2$& $1312.3 \pm 0.8$ & $935.3 \pm 0.8$ \\
 $\nu_3$& $1312.1 \pm 0.6$ &$955.9 \pm 0.7$ \\
 $\nu_4$& $1543.5 \pm 0.7$ &$1038.5 \pm 0.9$  \\
 $\nu_5$& $1543.0 \pm 0.7$ &$1224.3 \pm 0.9$ \\
 $\nu_6$& $2926.2 \pm 1.0$ & $1354.6 \pm 0.8$ \\
 $\nu_7$& $3038.3 \pm 1.0$ &$1445.0 \pm 1.3$  \\
 $\nu_8$& $3038.9 \pm 1.0$ &$1638.7 \pm 0.8$ \\
 $\nu_9$& $3039.0 \pm 0.7$ &$2977.6 \pm  1.0$ \\
 $\nu_{10}$& - &$2983.4 \pm 0.9$ \\
 $\nu_{11}$& - &$3054.9 \pm 0.8$\\
 $\nu_{12}$& - &$3083.5 \pm 0.8$\\
\noalign{\smallskip}\hline\noalign{\smallskip}
\end{tabular}
\end{table}

Next, we turn to the question of scaling. The cost is dominated by the PES
evaluation and is, therefore, measured by the number of samples $(S)$. The XVH2(4)
method is based on the truncated Taylor-series
expression of the PES and its force-constant evaluation is the hotspot of the whole
calculation. Its cost is exponential with the truncation rank $(q)$ and
high-rank ($q$th-order) polynomial with the number of modes (the molecular size,
$m$) thus resulting in $S = O(m^q)$. In our examples, $q = 4$ (quartic force
field) and by roughly doubling the molecular size, the cost of XVH2(4) increases
by a factor of $2^4 = 16$. Since the cost is exponential in $q$, the cost factor
is expected to be larger if higher order force constants are considered. In
CT-XVH2, the zero-point energies of water with $S = 150$ and those of
formaldehyde with $S = 4000$ seem to achieve comparable accuracy. Therefore the
cost increases by a factor of 27 upon doubling the molecular size. While the
absolute cost is still lower than XVH2, the dataset in this case is too small to
draw any definitive conclusion.

\begin{figure}[htb!]
\includegraphics[scale=.50]{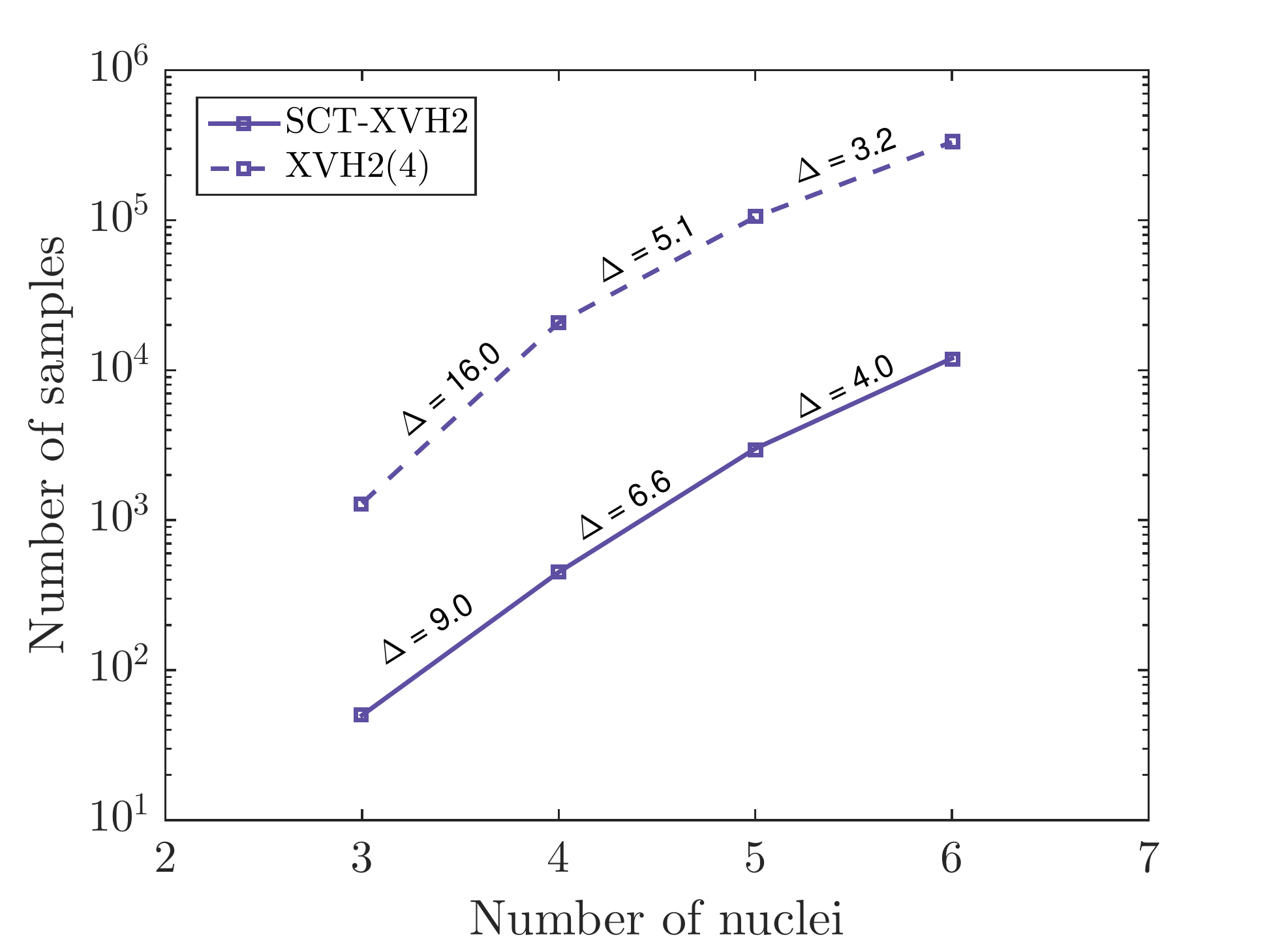}
\caption{Comparison of scaling of SCT-XVH2 and XVH2(4) methods. Cost factor $(\Delta)$ for successive molecular sizes in XVH2(4) is based on the relation $\Delta = \left(\frac{m_{a+1}}{m_a}\right)^q$, where $q = 4$ and $m_a = 3a-6$ is dimensionality of PES of a molecule with $a$ nuclei. Cost factor for SCT-XVH2 is the ratio of number of samples $(S)$ required to approximate PES of molecules with $(a+1)$ and $a$ nuclei for an accuracy $\epsilon_s \approx 1.0\times 10^{-2}$.}
\label{fig:scaling}
\end{figure}

With SCT-XVH2, we find that the cost increases by a factor of about 9 by
doubling the molecule size from water $(S = 50)$ to formaldehyde $(S = 450)$.
Similarly the cost factor increase is roughly 6.6 from formaldehyde to methane
$(S=3000)$ and reduces to 4 in case of methane to ethylene $(S=12000)$.
To compare scaling of SCT-XVH2 and XVH2(4), we define the cost factor $\Delta$ as the
ratio of the number of PES samples required to obtain zero point energies and
frequencies of molecules with $(a+1)$ and $a$ nuclei. Accordingly, in Figure
\ref{fig:scaling}, we plot scaling with cost factors of XVH2(4) such that
$\Delta = \left(\frac{m_{a+1}}{m_a}\right)^q$, where $q = 4$ and $m_a = 3a-6$ is
the dimensionality of the PES of a molecule with $a$ nuclei. Similarly, we calculate
$\Delta$ for SCT-XVH2 based on the number of samples $(S)$ required to
approximate the PES with an accuracy $\epsilon_s \approx 2.5\%$. We find
that, although SCT-XVH2 follows exponential scaling similar to XVH2(4), for the
molecules considered in this study, it has a smaller intercept implying a much
smaller absolute cost. Moreover, as in XVH2(4), the cost factor of SCT-XVH2 also
starts showing an apparent downward trend as we increase the number of atoms in the
system. For molecules bigger than those considered in this study, scaling of
SCT-XVH2 will depend on the sparsity of the PES on the chosen basis set.

\section{Conclusion}
\label{sec:conclusion}
We presented a general scalable approach that takes advantage of sparsity and
low rank structure to integrate high dimensional functions with minimal
evaluations of the integrand for a target accuracy. A sparse representation of
the integrand is sought after in an approximation space chosen \textit{a
priori}. The sparse solution thus obtained is then compressed using low rank
tensor decomposition to further reduce the number of terms in the separated
representation. Finally, an appropriate quadrature rule is used to perform
dimension-wise integration. We illustrated this method for approximating the PES
in XVH2 for calculating zero point energies and frequencies of molecules. The
method achieves similar accuracy, with orders of magnitude fewer evaluations, as
compared to existing methods in the literature.

In extending the proposed method beyond molecular
sizes considered in this work or to take into account effects of higher
quantum numbers, storage and compression of Green's function could become a
significant bottleneck. Also, currently we choose the basis set based on the
total degree of the multidimensional polynomial basis whose cardinality,
i.e. the number of bases, still has an exponential increase with
dimensionality of the PES. Moreover, selection of the basis using total
degree may exclude basis functions which could be significant for an
accurate sparse representation. In future work, we will enhance the approach
to overcome the current limitations. To deal with high dimensional Green's
function, we plan to use the recently proposed randomized CP-ALS algorithm
\cite{KOLDA17} where only a small subset of tensor entries are required to obtain 
a low rank decomposition, and which
can be evaluated on the fly. For better selection of basis set, we can use
adaptive basis selection based on interaction of atoms in the molecular
structure. Application of this
approach for larger molecules will lead to a better understanding of scaling
behavior.

\section{Acknowledgement} 
The authors thank Judit Zador at Sandia National Laboratories, Livermore for
fruitful discussions. Support for this work was provided through the Scientific
Discovery through Advanced Computing (SciDAC) program funded by the U.S.
Department of Energy, Office of Science, Advanced Scientific Computing Research
and Basic Energy Sciences under Award No.  DE-FG02-12ER46875. Sandia National
Laboratories is a multimission laboratory operated by National Technology and
Engineering Solutions of Sandia LLC, a wholly owned subsidiary of Honeywell
International Inc., for the U.S. Department of Energy's National Nuclear
Security Administration under contract DE-NA0003525. The views expressed in the article
do not necessarily represent the views of the U.S. Department of Energy or the
United States Government. Sandia has major research and development
responsibilities in nuclear deterrence, global security, defense, energy
technologies and economic competitiveness, with main facilities in Albuquerque,
New Mexico, and Livermore, California. This research used resources of the
National Energy Research Scientific Computing Center, a DOE Office of Science
User Facility supported by the Office of Science of the U.S.  Department of
Energy under Contract No. DE-AC02-05CH11231. 

\bibliography{paper}

\end{document}